\documentclass[fleqn,twoside]{article}
\usepackage{espcrc2}

\usepackage{amsmath}
\usepackage{graphicx}
\usepackage[figuresright]{rotating}

\title{Energy spectrum of a Heavy-Light meson}

\author{UKQCD Collaboration,
        J. Koponen\address[HY]{Department of Physical Sciences and
          Helsinki Institute of Physics,\\ 
          P.O. Box 64, FIN-00014 University of Helsinki, Finland},
          A. M. Green\addressmark[HY]
        and
        C. Michael\address{Department of Mathematical Sciences,\\
University of Liverpool, L69 3BX, UK}}
       
\begin{document}
\pagestyle{empty}
\begin{abstract}
The energies of different angular momentum states of a
heavy-light meson were measured on a lattice in \cite{MP}.
We have now repeated this study using a larger, quenched lattice
to check that there the finite size effects are under control.
Also the effect of quenching is studied by using dynamical
fermion configurations. In the latter case the
lattice spacing $a$ is smaller, which allows one to make an
estimate of $a\rightarrow 0$ corrections. The new measurements
confirm the earlier calculated energy spectrum.

From some theoretical considerations it is expected that, for
high angular momentum states, the multiplets should be
inverted compared with the Coulomb spectrum,
i.e. L$-$ should lie higher than L$+$ \cite{Schnitzer}.
This happens when a pure scalar confining potential is
present. Here the notation L$+$($-$) means that the light
quark spin couples to angular momentum L giving the total
$j=$L$\pm 1/2$. Experimentally this inversion is not seen
for P-wave mesons, and  lattice measurements now suggest that
there is no inversion in the D-wave either. Indeed we find
that the spin-orbit splitting is very small for both values
L$=1$ and $2$. This almost exact cancellation need not be a
coincidence, and could probably be understood by using
straightforward symmetry arguments \cite{Page}.
\vspace{1pc}
\end{abstract}

\maketitle

\section{Heavy-light mesons}
\label{hl}
A study of heavy-light mesons is not just for academic interest,
since they are also realised in nature. The best examples
are the $B(5.28~\textrm{GeV})$ and $B_s(5.37~\textrm{GeV})$ mesons,
which have quark structures $(\bar{b}u)$ and $(\bar{b}s)$. Since
the $b, \, s$ and $u$ quarks have masses of about $4.2, \, 0.1$ and
$0.001$~GeV, respectively, we see that the $B$ and $B_s$ are indeed
heavy-light mesons and can be thought of as the ``Hydrogen atom'' for
quark systems. Recently the discovery of a narrow resonance in the
$D_s\pi^0$ channel by the BABAR collaboration has provoked a lot of
interest in accurate measurements of the energies of higher angular
momentum states of mesons on a lattice. Unfortunately there is no
experimental data available on the P-wave state energies of $B$ or
$B_s$ mesons at present. The spectrum of $D_s$ mesons from lattice
is discussed for example in \cite{Bali,Dougall}.

\section{Lattice parameters}
In this study
the heavy quark is taken to be infinitely heavy, whereas the light
quark mass is approximately that of the strange quark. The lattice
parameters are given in Table \ref{param_table}.
\begin{table}[h!]
\centering
\caption{Lattice configurations used with Wilson glue and
tadpole-improved clover fermions. Dynamical configurations
(two quark flavours) have light valence quarks which are the
same as the sea-quarks \cite{allton}. The number
of configurations is 20 in the quenched cases ($N_f=0$) and
78 in the unquenched case ($N_f=2$). For $N_f=0$
$m_q/m_s\approx 0.83$ and in the $N_f=2$ case
$m_q/m_s\approx 1.28$. Here we take $r_0\approx 0.52$~fm.}
\label{param_table}
\begin{tabular}{cccccc}
\hline 
$\beta$ & $C_{SW}$ & Volume & $N_f$ & $\kappa$ & $r_0/a$\\
\hline  % or quote r_0 m_pi  or m_pi/m_rho or m_q/m_s 
 %no error on r_0 quenched
5.7 & 1.57 & $12^3 \times 24$ & 0 &0.14077 & 2.94 \\ 
5.7 & 1.57 & $16^3 \times 24$ & 0 &0.14077 & 2.94 \\ 
5.2 & 1.76 & $16^3 \times 24$ & 2 &0.1395  & 3.444\\ 
\hline
\end{tabular}
\end{table}

\section{Energy spectrum}
\subsection{Correlators}
\begin{figure}
\centering
  \includegraphics*[width=0.17\textwidth]{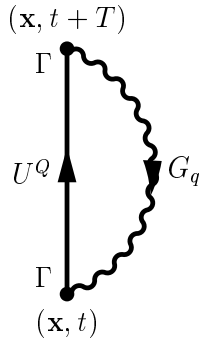}
\caption{The two-point correlation function $C_2$.}
\label{fig1}
\end{figure}

The basic quantity for evaluating the energies of heavy-light mesons is
the 2-point correlation function -- see Fig.~\ref{fig1}.
It is defined as
\begin{multline*}
C_2(T)=\langle P_t\Gamma G_q(\mathbf{x},t+T,t)\cdot \\
P_{t+T}\Gamma^{\dag}U^Q(\mathbf{x},t,t+T)\rangle,
\end{multline*}
where $U^Q(\mathbf{x},t,T)$ is the heavy (infinite mass)-quark propagator
and $G_q(\mathbf{x},t+T,t)$ the light anti-quark propagator. $P_t$
is a linear combination of products of gauge links at time $t$
along paths $P$ and $\Gamma$ defines the spin structure of the operator.
The $\langle ...\rangle$ means the average over the whole lattice.

\begin{figure}
\centering
\includegraphics*[angle=-90,width=0.46\textwidth]{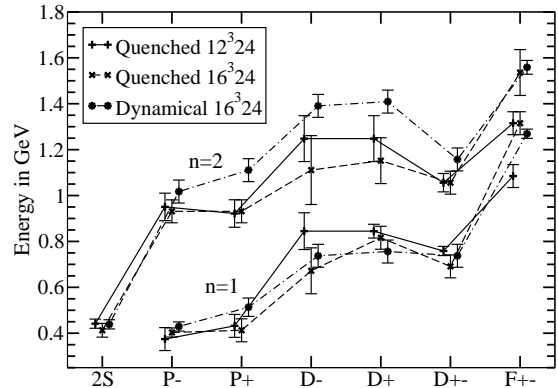}
\caption{The measured energy spectrum.}
\label{Espect}
\end{figure}

\subsection{Fits}
The energies are extracted by fitting the two-point
correlator $C_2$ with a sum of exponentials,
\begin{equation*}
C_2(T)\approx\sum_{i=1}^{N_i}c_{i}\mathrm{e}^{-m_i T},
\text{ where $N_i=2$ or $3$.}
\end{equation*}
The use of several fuzzing levels allows us to extract not only the lowest
mass but also the first excited state for different angular momentum states.
The resulting energy spectrum is shown in Fig. \ref{Espect}. When comparing
lattices of different sizes ($12^3$ vs. $16^3$) and different lattice spacings
($a=0.152$~fm vs. $a=0.18$~fm) a very good agreement is found. Effects coming
from quenching (compared with dynamical fermions) as well as finite size
effects and finite lattice spacing corrections seem to be small.

\section{Inversion}
\subsection{Theoretical expectations}
In the literature \cite{Schnitzer}, there have been claims that the
spectra of  heavy-light mesons should exhibit inverted multiplets,
where the L$+$ state lies lower in energy than the L$-$ state. For
small values of L --- in an interquark potential description --- the
one gluon exchange potential is expected to dominate leading to the
``natural'' ordering where the L$-$ state lies lower than the L$+$.
However, as L (or the principal quantum number n) increases, a model
containing a confining potential should eventually invert this
natural order. In \cite{Schnitzer} this claim is compared with
the experimentally observed 1P-states of the $D$ meson and the natural
ordering prevails.

\subsection{Dirac model}
\begin{figure}
\centering
\includegraphics*[angle=-90,width=0.46\textwidth]{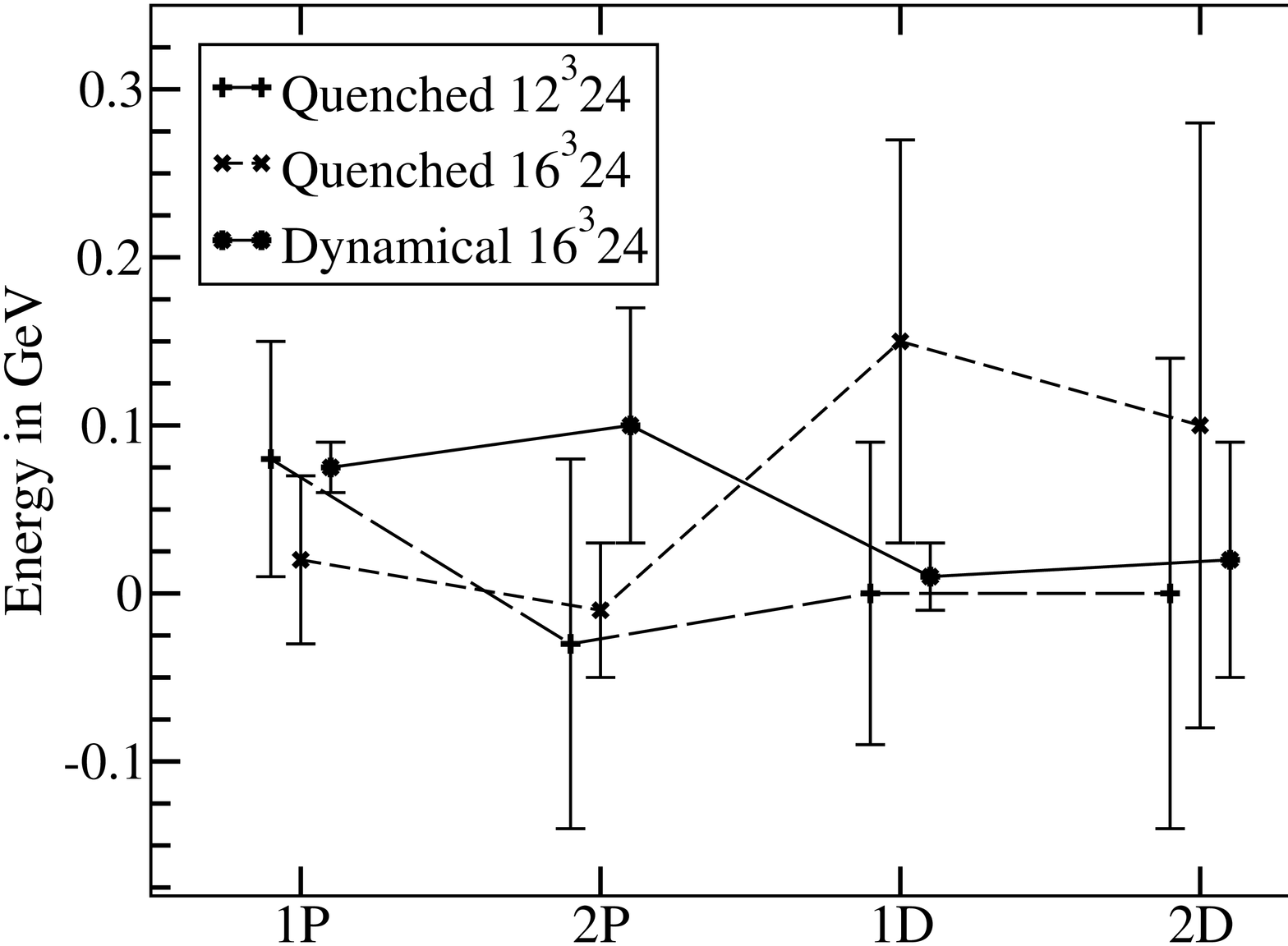}
\caption{The $E(\mathrm{L}+)-E(\mathrm{L}-)$ splitting.}
\label{fig3}
\end{figure}

In non-relativistic models of conventional me\-sons the splitting
between the L$+$ and L$-$ states is caused by introducing spin-orbit
interactions explicitly. In contrast, a Dirac equation description
with a spin-independent potential can automatically lead to spin-orbit
splitting. However, in the extreme case where the scalar ($V_s$)
and vector ($V_v$) potentials differ only by a constant $U$, i.e.
$V_v= V_s + U$, this splitting is naturally small \cite{Page}.
This could be approximately true asymptotically, i.e. when the distance
between the heavy quark and the light quark is large. In such a limit
a qualitative fit to our data can be obtained \cite{GIJK}. However,
Gromes seems to rule out vector exchange for the confining
quark-antiquark potential \cite{gromes}. The data
favour small $E(\mathrm{L}+)-E(\mathrm{L}-)$ splitting, as can be
seen in Fig.~\ref{fig3}.

\section{Conclusions and Future}
The results from three different lattices show that changing quenched
configurations to dynamical fermions does not lead to a large change
in energies. Also no signs of finite size effects or large errors due
to finite $a$ are observed. Spin-orbit splitting is found to be small
for both $P$- and $D$-wave states.

Much still remains to be done:
\begin{itemize}
\item Understand the energies and $S$-wave densities \cite{GKPM}
phenomenologically using the Dirac equation.

\item Measure correlations in  the baryonic and $(Q^2\bar{q}^2)$
systems \cite{TJ}.

\item Measure the $P$-, $D$- and $F$-wave densities.
\end{itemize}

The authors wish to thank the Center for Scientific Computing in Espoo,
Finland for making available resources without which this project could
not have been carried out. One of the authors (J.K.) wishes to thank the
Finnish Cultural Foundation and the Academy of Finland (project 177881)
for financial support.

\end{document}